\documentclass[pre,aps,amssymb,twocolumn,showpacs]{revtex4}
\usepackage[dvips]{graphicx}

\begin{document}

\title{Magnetic field inversion symmetry in quantum pumps with discrete symmetries}

\author{Sang Wook Kim}

\affiliation{Max-Planck-Institut f{\"u}r Physik komplexer Systeme, 
N{\"o}thnitzer Str. 38, 01187 Dresden, Germany}

\date{\today}

\begin{abstract}

We investigate the magnetic field inversion symmetry of the pumped currents in quantum 
pumps with various discrete symmetries using Floquet scattering matrix approach. 
We found the pumped currents can have symmetries $I(B,\phi) = -I(-B,-\phi)$ and 
$I(B,\phi) = I(-B,-\phi)$, where $\phi$ is the phase difference of two time dependent
perturbations, depending on the discrete symmetry considered. The results in the 
adiabatic limit for each discrete symmetry are compared with those of Brouwer's formula. 

\end{abstract}


\maketitle


\section{introduction}

The quantum pump is a device that generates a dc current at zero bias potential through
cyclic change of system parameters \cite{Hekking91,Altshuler99,Lubkin99}. Recently, 
adiabatic charge pumping in open quantum dots not only has attracted considerable 
theoretical attention 
\cite{Aleiner98,Brouwer98,Zhou99,Avron00,Shutenko00,Aleiner00,Wei00,Brouwer01,Vavilov01,Polianski01,Avron01,Sharma01,Entin-Wohlman02,Moskalets02a},
but was also experimentally realized by Switkes et al. \cite{Switkes99}.
After a cycle of the adiabatic shape change one returns to the initial configuration, but 
the wavefunction may have its phase changed with respect to the initial wavefunction. This 
is the geometric or Berry's phase \cite{Berry84}. The additional phase is equivalent 
to pumped charges that pass through the quantum dot \cite{Altshuler99}. 
From another point of view, the quantum pump is a time dependent system driven by (at least) 
two different time periodic perturbations with the same angular frequency and a phase 
difference $\phi$. One can deal with this problem under the adiabatic approximation, or 
by using the more general Floquet approach \cite{Sambe73,Kim02_qp,Moskalets02b}. Recently, 
it has been shown that 
in the adiabatic limit with small strength of the oscillation potentials the Floquet and the 
adiabatic approach give exactly equivalent results \cite{Kim02_qp,Moskalets02b}.

The puzzle unsolved in the experiment performed by Switkes et al. \cite{Switkes99} is the 
magnetic field inversion symmetry (MIS) in pumped currents. It was suggested theoretically 
\cite{Zhou99} that the pumped current $I$ is invariant upon magnetic field reversal
\begin{equation}
I(B) = I(-B).
\label{symmetry}
\end{equation}
The subsequent experiment on open quantum dots appears to be in a good agreement with 
Eq.~(\ref{symmetry}) \cite{Switkes99}. It became immediately clear, however, that such 
symmetry is not valid in theory \cite{Shutenko00}. Recently, Brouwer proposed that 
the rectified displacement current could account for the MIS \cite{Brouwer01}, which means 
the current observed in the experiment is not attributed to the adiabatic quantum pumping.

Even though in general quantum pumps do not fulfill the MIS, additional discrete symmetries 
of quantum dots can lead to such MIS. The effect of discrete symmetries on the MIS has been 
studied by Aleiner, Altshuler, and Kamenev \cite{Aleiner00} in the adiabatic limit. They found 
the reflection symmetries give rise to relations $I(B)=I(-B)$ or $I(B)=-I(-B)$ depending on 
the orientation of the reflection axis. In the presense of inversion $I(B)=0$. However, the 
symmetry considered in Ref. \cite{Aleiner00} should be kept intact in the pumping cycle. 
The only way to realize such symmetry is to set $\phi=n\pi$, but then there is no adiabatic 
pumped current. 

\begin{figure}
\center
\includegraphics[height=12 cm,angle=0]{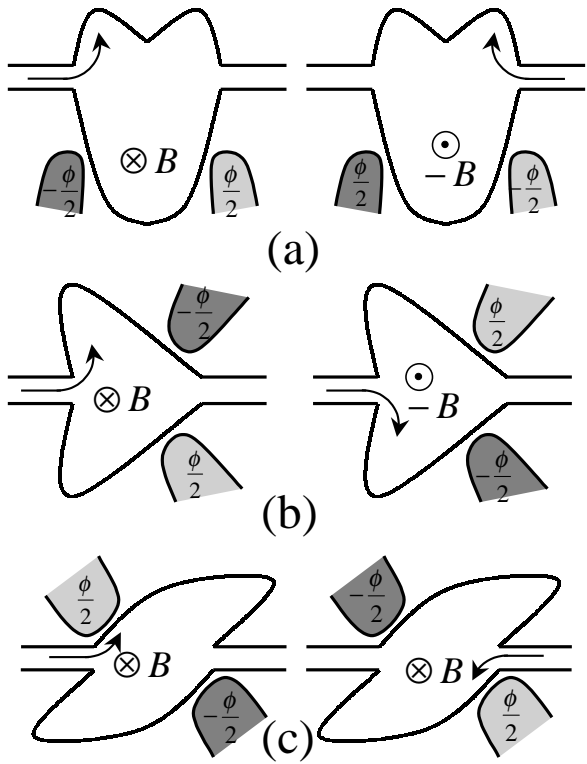}
\caption{The schematic diagram for three discrete symmetries of quantum pumps, i.e. (a) LR, 
(b) UD and (c) IV symmetry.}
\label{fig1}
\end{figure}

In this paper, by using Floquet scattering matrix approach we investigate the MIS of the pumped 
current for three discrete symmetries, namely LR (left-right), UD (up-down), and IV (inversion), 
as shown in Fig.~\ref{fig1}, both in the adiabatic and in the non-adiabatic cases. Since these 
symmetries are only applied to the {\em time independent} part of the scattering potential with 
an arbitrary $\phi$ they are more relevant for applying to real experimental situation. 

For LR symmetry, we find
\begin{equation}
I_{LR}(B,\phi)=-I_{LR}(-B,-\phi)
\label{LRsym}
\end{equation}
both in the adiabatic and in the non-adiabatic cases. On the other hand, dots with UD
symmetry obey
\begin{equation}
I_{UD}(B,\phi)=I_{UD}(-B,-\phi)
\label{UDsym}
\end{equation}
both in the adiabatic and in the non-adiabatic cases, as well. IV symmetry does not
have any relevant symmetry for the magnetic field inversion. From the Floquet scattering 
approach, however, the pumped current is shown to be vanishingly small in the adiabatic limt. 
It is emphasized that Eqs.~(\ref{LRsym}) and (\ref{UDsym}) do not correspond to 
$I_{LR}(B) = I_{LR}(-B)$ and $I_{UD}(B) = -I_{UD}(-B)$, respectively. In the adiabatic 
limit it is always true that $I(B,\phi)=-I(B,-\phi)$, consequently $I(B,\phi=0)=0$, while 
it is no longer available in the non-adiabatic case.

In Sec.~II Floquet scattering matrix formalism is presented for the case without magnetic
field. In Sec.~III we discuss magnetic field inversion operation in a quantum pump, and
investigate the MIS for three discrete symmetries, i.e. LR, UD, and IV symmetry. In 
Sec.~IV we compare the results obtained in Sec.~III with those from Brouwer's formula, 
which shows they are consistent with each other in the adiabatic limit. Finally, we 
conclude in Sec.~V.


\section{Floquet scattering matrix formalism for a quantum pump}

First, we introduce Floquet scattering matrix formulation without magnetic field for
simplicity. Consider the one-dimensional time-dependent Schr\"odinger equation 
$i\hbar(\partial/\partial t)\psi = H(t)\psi$ for a non-interacting electron 
with mass $\mu$ and $H(t) = -\hbar^2\nabla^2/2\mu + U(x,t)$, where $U(x,t+T)=U(x,t)$ and 
$U(x,t)=0$ at $x\rightarrow \pm\infty$. Due to the periodicity in time, a solution
can be written as
\begin{equation}
\Psi_\epsilon(x,t) = e^{-i \epsilon t/\hbar}\sum_{n=-\infty}^{\infty} \chi_n (x) e^{-in\omega t},
\label{floquet}
\end{equation}
where $\epsilon$ is the Floquet energy which takes a continuous value in the interval
$[0,\hbar \omega)$. Since the potential is zero at $x\rightarrow \pm \infty$, 
$\chi_n(x)$ is given by the following form
\begin{equation}
\chi_n (x) = \left\{
\begin{array}{c}
A_n e^{ik_n x} + B_n e^{-ik_n x}, ~~~ x\rightarrow -\infty \\
C_n e^{ik_n x} + D_n e^{-ik_n x}, ~~~ x\rightarrow +\infty,
\end{array} \right.
\label{plane_wave}
\end{equation}
where $k_n=\sqrt{2\mu(\epsilon + n\hbar\omega)}/\hbar$. By wave matching, the incoming and the 
outgoing waves can be connected by matrix $M$
\begin{equation}
\left(
        \begin{array}{c}
        \vec{B} \\ \vec{C}
        \end{array}
\right)
=M
\left(
        \begin{array}{c}
        \vec{A} \\ \vec{D}
        \end{array}
\right).
\label{mat_eq}
\end{equation}
If we keep only the propagating modes ($k_n$ is real), we can obtain the unitary Floquet 
scattering matrix \cite{unitarity}, which can be expressed in the following form 
\cite{Li99,Henseler01,Kim02_pole}
\begin{equation}
S = \left(
        \begin{array}{cccccc}
        r_{00} & r_{01} & \cdots & t'_{00} & t'_{01} & \cdots \\
        r_{10} & r_{11} & \cdots & t'_{10} & t'_{11} & \cdots \\
        \vdots & \vdots & \ddots & \vdots  & \vdots  & \ddots \\
        t_{00} & t_{01} & \cdots & r'_{00} & r'_{01} & \cdots \\
        t_{10} & t_{11} & \cdots & r'_{10} & r'_{11} & \cdots \\
        \vdots & \vdots & \ddots & \vdots  & \vdots  & \ddots \\
        \end{array}
\right),
\label{osc_smatrix}
\end{equation}
where $r_{nm}$ and $t_{nm}$ are the reflection and the transmission amplitudes respectively,
for modes incident from the left with energy $\epsilon+m\hbar\omega$ and outgoing to modes
with energy $\epsilon+n\hbar\omega$; $r'_{nm}$ and $t'_{nm}$ are similar quantities for modes
incident from the right. 

The total transmission coefficient from the left to the right as a function of an energy of 
an incident electron is given by
\begin{equation}
T^{\rightarrow}(E)=\sum_{n=0}^{\infty}\left| t_{nm} (\epsilon) \right|^2,
\label{Fl_trans}
\end{equation}
where $E=\epsilon+m\hbar\omega$. The total transmission from the right to the left 
$T^\leftarrow(E)$ can also be determined in a similar way. Then, the pumped current to 
the right $I_r$ is given by the difference between two currents having the opposite directions
\cite{Kim02_qp,Moskalets02b}:
\begin{equation}
I_r = I^\rightarrow - I^\leftarrow,
\label{floq_p_current}
\end{equation}
where
\begin{equation}
I^{\rightarrow(\leftarrow)} = \frac{2e}{h}\int_0^\infty dE f(E) T^{\rightarrow(\leftarrow)}(E).
\label{pump_current}
\end{equation}
Here, $f(E)$ represents the Fermi-Dirac distribution. Without a bias $f(E)$ has the same form
for all reservoirs. The pumped current to the left $I_l$ is equal to $-I_r$. Without specification
$I$ denotes $I_r$ below. To derive Eq.~(\ref{pump_current}), we need to assume that the reservoirs
are always at equilibrium even under external time-dependent perturbation \cite{Kim02_pauli}. 
This is guaranteed if the relaxation time of an electron in the reservoirs is much faster than 
a pumping cycle $T$.


\section{magnetic field inversion in a quantum pump}

\begin{figure}
\center
\includegraphics[height=6.5 cm,angle=0]{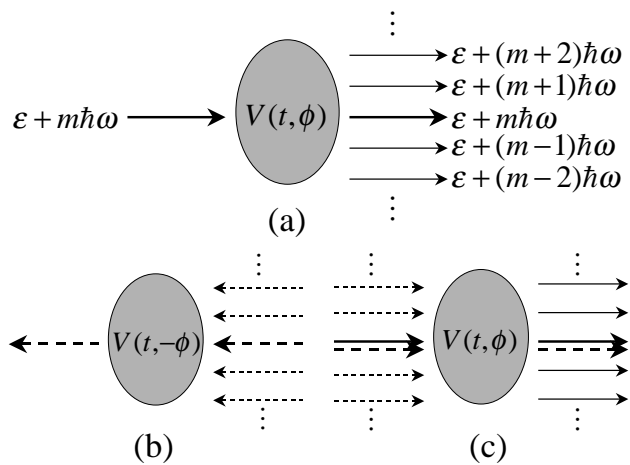}
\caption{Schematic diagrams for describing the configuration of side bands with given channels 
$\alpha$ and $\beta$ to calculate the total transmission coefficients through the time-periodic 
perturbation $V(t)$ for (a) $T^\rightarrow_{B,\phi} (E)$ and (b) the case given in 
Eq.~(\ref{eq-b}). (c) The setup (a) (the solid arrows) superimposed by the setup (b) reversed 
by using IV symmetry given in Eq.~(\ref{IVsym_coef}) (the dashed arrows).}
\label{fig2}
\end{figure}

Let us now consider the two-dimensional quantum dot driven by time-periodic perturbations 
under a static magnetic field, which is attached to two leads (see Fig.~\ref{fig2}), 
whose Hamiltonian is given by
\begin{equation}
i\hbar \frac{\partial\psi}{\partial t} 
= \left[\frac{(i\hbar\nabla + eA)^2}{2\mu} +V({\bf r},t)\right]\psi,
\label{schro_b}
\end{equation}
where $A$ is the vector potential. The potential $V({\bf r},t)$ is given by 
\begin{equation}
V({\bf r},t,\phi) = V_0({\bf r}) 
+ V_1({\bf r}) \cos(\omega t-\phi/2) + V_2({\bf r}) \cos(\omega t +\phi/2), 
\label{potential}
\end{equation}
where $V_0$ represents the confined potential of a quantum dot, and $V_1$ and $V_2$ are
the spatial dependence of two time dependent perturbation. $\phi$ is the initial phase 
difference of two time-dependent perturbations, and ${\bf r}$ denotes $(x,y)$. Note that 
$V({\bf r},-t,\phi)=V({\bf r},t,-\phi)$. When $\phi=n\pi$ without magnetic field, the 
quantum pump is microscopically reversible (or time reversible), i.e. 
$t^{\beta\alpha}_{nm} = t'^{\alpha\beta}_{mn}$, where $t_{nm}^{\beta\alpha}$ represents 
the transmission amplitude from the Floquet side band $m$ of the channel $\alpha$ in the 
left lead to the side band $n$ of the channel $\beta$ in the right lead, and 
$t'^{\alpha\beta}_{mn}$ is a similar quantity for the opposite direction. In a quantum
pump the two time dependent perturbations with a finite $\phi$ ($\neq n\pi$) as well 
as the magnetic field break the time reversal symmetry. 

If we take the complex conjugate of Eq.~(\ref{schro_b}) and at the same time, reverse 
the vector potential ($A \rightarrow -A$) and time ($t \rightarrow -t$), we obtain
\begin{equation}
i\hbar \frac{\partial\psi^*}{\partial t} 
= \left[\frac{(i\hbar\nabla + eA)^2}{2\mu} +V({\bf r},-t)\right]\psi^*.
\end{equation}
Compared with Eq.~(\ref{schro_b}) one can see that 
$\psi_{-B,\phi}({\bf r},t) =\psi^*_{B,-\phi}({\bf r},-t)$. 
The complex conjugation combined with the time inversion of the solution in Eqs.~(\ref{floquet}) 
and (\ref{plane_wave}) simply corresponds to the reversal of the momentum direction of the 
incoming or the outgoing plane waves and simultaneously the replacement of absorption by 
emission, and vice versa. This simply implies that $S \rightarrow S^T$ ($S^T$ is the transpose 
of $S$), i.e. $S_{nm}^{\beta\alpha} \rightarrow S_{mn}^{\alpha\beta}$. 
For completion of the magnetic field inversion operation, $\phi \rightarrow -\phi$ also has to 
be considered.

The total transmission coefficient to the right reservoir after the magnetic field inversion 
is given by
\begin{eqnarray}
T^\rightarrow_{-B,\phi}(E) & = & 
\sum_{\alpha\beta}\sum_{n=0}^{\infty}\left| t_{nm}^{\beta\alpha}(\epsilon,-B,\phi)\right|^2 
\label{eq-a}\\
& = & \sum_{\alpha\beta}\sum_{n=0}^{\infty}
\left| t'^{\alpha\beta}_{mn}(\epsilon,B,-\phi)\right|^2, \label{eq-b}
\end{eqnarray}
where $E=\epsilon+m\hbar\omega$. It must be noted that Eq.~(\ref{eq-b}) is not equivalent to 
$T^\leftarrow_{B,-\phi}$. Figures \ref{fig1}(a) and (b) represent the usual setup of side 
bands for calculating the total transmission coefficients, e.g. $T^\rightarrow_{B,\phi}(E)$, 
and the setup for calculating Eq.~(\ref{eq-b}), respectively. In general, they are different 
from each other, so are $I(B)$ and $I(-B)$. The MIS is not valid in Floquet formalism, as
well.


\subsection{LR symmetry}

If we assume LR symmetry [$V_0(x,y)=V_0(-x,y)$ and $V_1(x,y)=V_2(-x,y)$], the following 
relations are obtained [see Fig.~\ref{fig1}(a)] 
\begin{eqnarray}
t^{\beta\alpha}_{nm}(\epsilon,B,\phi) & = & t'^{\beta\alpha}_{nm}(\epsilon,-B,-\phi), \nonumber \\
r^{\beta\alpha}_{nm}(\epsilon,B,\phi) & = & r'^{\beta\alpha}_{nm}(\epsilon,-B,-\phi).
\label{LRsym_coef}
\end{eqnarray}
Then, Eq.~(\ref{eq-a}) can be rewritten as
\begin{eqnarray}
T^\rightarrow_{-B,\phi}(E) & = & \sum_{\alpha\beta}\sum_{n=0}^{\infty}
\left| t'^{\beta\alpha}_{nm}(\epsilon,B,-\phi)\right|^2 \nonumber \\
& = & T^\leftarrow_{B,-\phi}(E).
\end{eqnarray}
From Eq.~(\ref{floq_p_current}) one reaches $I(B,\phi)=-I(-B,-\phi)$, i.e. Eq.~(\ref{LRsym}). 
Since such relation results from the Floquet approach, it is available for both the adiabatic 
and the non-adiabatic cases. Note that when $\phi=0$ the result obtained in Ref. \cite{Aleiner00} 
is recovered, and its validity is now extended even to the non-adiabatic case. At zero magnetic 
field one finds $I(\phi)=-I(-\phi)$, consequently $I(\phi=0)=0$ even for the non-adiabatic case.


\subsection{UD symmetry}

For UD symmetry [$V_0(x,y)=V_0(x,-y)$ and $V_1(x,y)=V_2(x,-y)$] one finds
\begin{eqnarray}
t^{\beta\alpha}_{nm}(\epsilon,B,\phi) & = & t^{\beta\alpha}_{nm}(\epsilon,-B,-\phi), \nonumber \\
r^{\beta\alpha}_{nm}(\epsilon,B,\phi) & = & r^{\beta\alpha}_{nm}(\epsilon,-B,-\phi)
\label{UDsym_coef}
\end{eqnarray}
[see Fig.~\ref{fig1}(b)]. Then, Eq.~(\ref{eq-a}) can be rewritten as
\begin{eqnarray}
T^\rightarrow_{-B,\phi}(E) & = & \sum_{\alpha\beta}\sum_{n=0}^{\infty}
\left| t_{nm}^{\beta\alpha}(\epsilon,B,-\phi)\right|^2 \nonumber \\
& = & T^\rightarrow_{B,-\phi}(E),
\end{eqnarray}
which leads to $I(B,\phi)=I(-B,-\phi)$, i.e. Eq.~(\ref{UDsym}). Like the LR symmetry, such 
symmetry of the quantum pump is correct for both the adiabatic and the non-adiabatic cases.
Note also that when $\phi=0$ the result obtained in Ref. \cite{Aleiner00} is recovered
not only for the adiabatic limit but also for the non-adiabatic case. At zero magnetic 
field one finds $I(\phi)=I(-\phi)$, which can be understood from the fact that the exchange 
of $V_1$ and $V_2$, i.e. $\phi \rightarrow -\phi$, does not make any difference in 
Fig.~\ref{fig1}(b).


\subsection{IV symmetry}

Finally, we consider IV symmetry [$V_0(x,y)=V_0(-x,-y)$ and $V_1(x,y)=V_2(-x,-y)$]. 
If one assume IV symmetry, one obtains [see Fig.~\ref{fig1}(c)]
\begin{eqnarray}
t^{\beta\alpha}_{nm}(\epsilon,B,\phi) = t'^{\beta\alpha}_{nm}(\epsilon,B,-\phi), \nonumber \\
r^{\beta\alpha}_{nm}(\epsilon,B,\phi) = r'^{\beta\alpha}_{nm}(\epsilon,B,-\phi).
\label{IVsym_coef}
\end{eqnarray}
Eq.~(\ref{eq-b}) is reexpressed as
\begin{equation}
T^\rightarrow_{-B,\phi}(E) = \sum_{\alpha\beta}\sum_{n=0}^{\infty}
\left| t_{mn}^{\alpha\beta}(\epsilon,B,\phi)\right|^2,
\end{equation}
which is different from $T^\rightarrow_{B,\phi}(E)$ since the summation is taken for $n$
as shown in Fig.~\ref{fig2}(c). In general, there is no relevant MIS for dots with IV symmetry. 
When $\phi=0$, however, Eq.~(\ref{IVsym_coef}) leads to $T^\rightarrow_B(E)=T^\leftarrow_B(E)$,
consequently $I=0$. The result in Ref.~\cite{Aleiner00} is also recovered for both the adiabatic
and non-adiabatic case.


\section{adiabatic limit}

The adiabatic condition in the quantum pump implies that any time scale of the problem 
considered must be much smaller than the period of the oscillation of an external pumping 
\cite{Brouwer98}. We can then define the instantaneous scattering matrix with time dependent 
parameters, namely $X_n(t)$,
\begin{equation}
\hat{S}_{ad}(E,t) = \hat{S}_{ad}(E,\{X_n(t)\})=
\left(
        \begin{array}{cc}
        \hat{r}_{ad} & \hat{t}'_{ad} \\ 
        \hat{t}_{ad} & \hat{r}'_{ad} 
        \end{array}
\right)
\label{adiabatic}
\end{equation}
Due to the time periodicity of $X_n$'s, using a Fourier transform one can obtain the amplitudes
of side bands for particles traversing the adiabatically oscillating scatterer with incident
energy $E$ as following
\begin{equation}
\hat{S}_{ad}(E,\{X_n(t)\}) = \sum_n \hat{S}_{ad,n}(E)e^{-in\omega t},
\label{adiabatic1}
\end{equation}
where
\begin{equation}
\hat{S}_{ad,n}(E)=\frac{1}{T}\int_0^T dt~e^{in\omega t} \hat{S}_{ad}(E,\{X_n(t)\}).
\label{adiabatic2}
\end{equation}
Thus we can construct the adiabatic Floquet scattering matrix as follows \cite{Moskalets02b}
\begin{equation}
S(E_n,E) \approx S(E,E_{-n}) \equiv \hat{S}_{ad,n}(E),
\label{moska_adia}
\end{equation}
where $E_n=E+n\hbar\omega$. By exploiting Eqs.~(\ref{adiabatic1}), (\ref{adiabatic2}) and 
(\ref{moska_adia}) the MIS's of adiabatic scattering matrices can be obtained directly from 
those of Floquet scattering matrices for the three symmetry classes.

Once the adiabatic scattering matrix is found, by using Brouwer's formula \cite{Brouwer98} 
the pumped charge $Q$ per one cycle can be expressed as
\begin{equation}
Q(B)=g[\hat{t}_{ad}(B),\hat{r}'_{ad}(B)] \equiv \int^T_0 dt \{f[\hat{t}_{ad}(B)]+f[\hat{r}'_{ad}(B)]\},
\label{Brouwer_eq}
\end{equation}
where
\begin{equation}
f(s)=\frac{e}{2\pi}\sum^2_{k=1}\sum_{\alpha\beta} {\rm Im}
\frac{\partial s_{\alpha\beta}}{\partial X_k}s^*_{\alpha\beta} \frac{dX_k}{dt}.
\end{equation}
Then, the pumped current $I$ is $Q/T$. Below, we will use Eq.~(\ref{Brouwer_eq}) for the
current.


\subsection{LR symmetry}

The relation (\ref{LRsym_coef}) implies $\hat{t}(\hat{r})^{\beta\alpha}_{ad,n}(E;B,\phi)
=\hat{t}'(\hat{r}')^{\beta\alpha}_{ad,n}(E;-B,-\phi)$ in the adiabatic limit, which, by using 
Eq.~(\ref{adiabatic1}), immediately leads to $\hat{t}(\hat{r})_{ad}(E,t;B,\phi)=
\hat{t}'(\hat{r}')_{ad}(E,t;-B,-\phi)$. From Eq.~(\ref{Brouwer_eq}) one finds 
\begin{eqnarray}
I_r(-B,\phi) & = & g[\hat{t}(-B,\phi),\hat{r}'(-B,\phi)] \nonumber \\
& = & g[\hat{t}'(B,-\phi),\hat{r}(B,-\phi)] \\
& = & I_l(B,-\phi) = -I_r(B,-\phi). \nonumber
\end{eqnarray}
This exactly corresponds to Eq.~(\ref{LRsym}). Note that if the relation $I(\phi)=-I(-\phi)$,
which is available only in the adiabatic limit, is considered, one reaches simpler form 
$I(B)=I(-B)$.


\subsection{UD symmetry}

Like the previous subsection the relation (\ref{UDsym_coef}) implies 
$\hat{t}(\hat{r})^{\beta\alpha}_{ad,n}(E;B,\phi)=\hat{t}(\hat{r})^{\beta\alpha}_{ad,n}(E;-B,-\phi)$ 
in the adiabatic limit, which, by using Eq.~(\ref{adiabatic1}), leads to 
$\hat{t}(\hat{r})_{ad}(E,t;B,\phi)=\hat{t}(\hat{r})_{ad}(E,t;-B,-\phi)$. From Eq.~(\ref{Brouwer_eq}) 
one obtains 
\begin{eqnarray}
I_r(-B,\phi) & = & g[\hat{t}(-B,\phi),\hat{r}'(-B,\phi)] \nonumber \\
& = & g[\hat{t}(B,-\phi),\hat{r}'(B,-\phi)] \\
& = & I_r(B,-\phi). \nonumber
\end{eqnarray}
This exactly corresponds to Eq.~(\ref{UDsym}). Note also that if the relation $I(\phi)=-I(-\phi)$
is considered, one reaches simpler form $I(B)=-I(-B)$.


\subsection{IV symmetry}

The relation (\ref{IVsym_coef}) implies $\hat{t}(\hat{r})^{\beta\alpha}_{ad,n}(E;B,\phi)
=\hat{t}'(\hat{r}')^{\beta\alpha}_{ad,n}(E;B,-\phi)$ in the adiabatic limit, which, by using 
Eq.~(\ref{adiabatic1}), leads to $\hat{t}(\hat{r})_{ad}(E,t;B,\phi)=
\hat{t}'(\hat{r}')_{ad}(E,t;B,-\phi)$. Considering the magnetic field inversion operation, 
one can obtain 
\begin{eqnarray}
\hat{t}_{ad}(E,t;B,\phi) & = & \hat{t}^T_{ad}(E,t;-B,\phi), \nonumber \\
\hat{r}_{ad}(E,t;B,\phi) & = & \hat{r}'^T_{ad}(E,t;-B,\phi). 
\label{adia_sym}
\end{eqnarray}
From Eq.~(\ref{Brouwer_eq}) one finds 
\begin{eqnarray}
I_r(-B,\phi) & = & g[\hat{t}(-B,\phi),\hat{r}'(-B,\phi)] \nonumber \\
& = & g[\hat{t}^T(B,\phi),\hat{r}^T(B,\phi)]. 
\end{eqnarray}
There is no relevant symmetry in the adiabatic limit, again. 

Eq.~(\ref{moska_adia}) implies that $t_{m+n,m} \approx t_{m,m-n}$ in the adiabatic limit. 
After summing up all the side band contributions, the two configurations shown in 
Fig.~\ref{fig1}(c) (the solid and the dashed arrows) give the approximately equivalent total 
transmission coefficients, i.e. 
$T^\rightarrow_{B}(E) \approx T^\rightarrow_{-B}(E)$. By applying similar procedure to the 
reflection coefficient one can also find $R^\rightarrow_{B}(E) \approx R^\leftarrow_{-B}(E)$ 
(make sure that the arrow is now reversed). It is also satisfied that 
$T^\rightarrow_{B}(E) \approx T^\leftarrow_{-B}(E)$ since the unitarity of Floquet scattering 
matrix requires $T^{\rightarrow(\leftarrow)}+R^{\rightarrow(\leftarrow)}=1$. From these two
relations the pumped current is shown to be vanishingly small, 
$I(B)=\int dE [T^\rightarrow_B(E) -T^\leftarrow_B (E)] \approx 0$ since 
$T^\rightarrow_{B}(E) \approx T^\rightarrow_{-B}(E) \approx T^\leftarrow_{B}(E)$.
In Table~I, we summarize the MIS of the pumped currents with various situations for the
three symmetries.


\section{summary and discussion}

In summary, we have investigated magnetic field inversion symmetry in a quantum pump
with various discrete symmetries using Floquet scattering matrix approach. We found the 
pumped currents can possess symmetries $I(B,\phi) = -I(-B,-\phi)$ for LR symmetry, and
$I(B,\phi) = I(-B,-\phi)$ for UD symmetry. For IV symmetry there is no relevant symmetry.
The adiabatic limit for each discrete symmetry has been considered by using Brouwer's 
formula. 

One possibility to realize the experimentally observed MIS, i.e. $I(B)=I(-B)$, is that
the quantum dot would obey LR symmetry. Then, the MIS is recovered only {\em in the
adiabatic limit}, while it will be broken in non-adiabatic case. In general, however, 
it is difficult that a chaotic quantum dot possesses any discrete symmetry like the LR 
symmetry.

\section*{Acknowledgments}
I would like to thank Michael Moskalets, Henning Schomerus, and Hwa-Kyun Park for 
useful discussions.


\bibliographystyle{prsty}

\begin{thebibliography}{10}
\bibitem{unitarity}
Eq.~(\ref{mat_eq}) can also be expressed in the form 
$\left.|{\rm out}\right> = M \left.|{\rm in}\right>$. In order to construct the 
scattering matrix we multiply an identity to both sides, 
$K^{-1}K\left.|{\rm out}\right> = M K^{-1}K\left.|{\rm in}\right>$, where $K_{nm}=\sqrt{k_n}
\delta_{nm}$. Then we have $\vec{J}_{out}=\bar{M}\vec{J}_{in}$, where $\vec{J}$ represents
the amplitude of probability flux and $\bar{M} \equiv KMK^{-1}$. Eq.~(\ref{osc_smatrix}) is
obtained from $\bar{M}$.
\end{thebibliography}

\begin{table}
\caption{The symmetries of the pumped currents for the magnetic field inversion.
TRS denotes time reversal symmetry by two time dependent perturbations, i.e.
TRS for $\phi=n\pi$ and no TRS for $\phi \neq n\pi$. $\times$ represents that
there is no relevant symmetry.}
\begin{tabular}{ccccc}
\hline
 & \multicolumn{2}{c}{adiabatic} & \multicolumn{2}{c}{non-adiabatic} \\
 & TRS & no TRS & TRS & no TRS \\ \hline
LR & $I=0$ & $I(B)=I(-B)$ & $I(B)=-I(-B)$ & $I(B,\phi)=-I(-B,-\phi)$ \\
UD & $I=0$ & $I(B)=-I(-B)$ & $I(B)=I(-B)$ &  $I(B,\phi)=I(-B,-\phi)$ \\
IV & $I=0$ & $I\approx 0$ & $I=0$ &  $\times$ \\
\hline
\end{tabular}
\end{table}

\end{document}